\documentclass[11pt,a4paper]{article}
\usepackage{jcappub}
\usepackage{bm}
\usepackage{dcolumn}
\usepackage{graphicx}

\newif\ifAMStwofonts
\AMStwofontstrue

\def\gsim{~\rlap{$>$}{\lower 1.0ex\hbox{$\sim$}}}

\def\simpropto{\lower.2ex\hbox{$\; \buildrel \propto \over \sim \;$}}
\def\ltsim{\lower.5ex\hbox{$\; \buildrel < \over \sim \;$}}
\def\gtsim{\lower.5ex\hbox{$\; \buildrel > \over \sim \;$}}
\def\ltsim{\lower.5ex\hbox{$\; \buildrel < \over \sim \;$}}
\def\gtsim{\lower.5ex\hbox{$\; \buildrel > \over \sim \;$}}

\def\kms{\mbox{km\,s$^{-1}$}}

\def\dd{\,{\rm d}}

\def\kms{\ {\rm km\,s^{-1}}}
\def\hmpc{\ {\rm h^{-1}Mpc}}

\def\dd{{\rm d}}

\def\grad{\nabla}
\def\pmb#1{\setbox0=\hbox{#1}%
\kern-.025em\copy0\kern-\wd0
\kern.05em\copy0\kern-\wd0
\kern-.025em\raise.0433em\box0}

\def\vv{\boldsymbol{v}}

\def\vY{\boldsymbol{Y}}
\def\vpsi{\boldsymbol{\psi}}

\newcommand{\sspa}{$\mathbf{s}$-space}

\def\vs{\boldsymbol{s}}

\def\vx{\boldsymbol{x}}
\def\vy{\boldsymbol{y}}
\def\vr{\boldsymbol{r}}

\def\vu{\boldsymbol{u}}
\def\va{{\bf a}}
\def\hvn{\hat {\vx}}
\def\hvx{\hat {\vx}}
\def\hvy{\hat {\vy}}

\def\hva{\hat {\va}}

\def\vk{\boldsymbol{k}}
\def\hvk{\hat {\vk}}

\def\simlt{\lower.5ex\hbox{$\; \buildrel < \over \sim \;$}}
\def\simgt{\lower.5ex\hbox{$\; \buildrel > \over \sim \;$}}

\newcommand{\beq}{\begin{equation}}
\newcommand{\eeq}{\end{equation}}
\def\beqa{\begin{eqnarray}}
\def\eeqa{\end{eqnarray}}
\def\fixit#1{}
\def\hmpc{h^{-1}\,{\rm Mpc}}

\def\dd{{\rm d}}

\notoc

\begin{document}

\title{Beyond boundaries of redshift surveys: assessing mass fluctuations on ``super-survey" scales}

\author{Martin Feix}
\author{and Adi Nusser}

\affiliation{Department of Physics and the Asher Space Science Institute,\\ Israel Institute of Technology - Technion, Haifa 32000, Israel}

\emailAdd{mfeix@physics.technion.ac.il}
\emailAdd{adi@physics.technion.ac.il}

\abstract{The observed density field in redshift space is directly affected by the radial motions generated from mass fluctuations outside the volume
occupied by a given galaxy redshift survey. These motions introduce redshift space anisotropies which are more pronounced at larger distances from the
survey's center, thus offering clues to the nature of mass fluctuations on super-survey scales. Furthermore, we note that all estimates of the growth
factor derived from redshift space distortions are based on relations which explicitly assume that the velocity field is generated by mass fluctuations
inside the survey volume. This may cause uncertainties in these estimates which are on the order of a few percent.}


\maketitle

\section{Introduction}
\label{sec:int}

The backbone of the Big Bang scenario is the cosmological principle which we interpret here as the approach to homogeneity on the
largest physical scales.\footnote{Einstein phrased the cosmological principle as \emph{Alle Stellen des Universums sind gleichwertig;
im speziellen soll also auch die \"{o}rtlich gemittelte Dichte der Sternmaterie \"{u}berall gleich sein} \cite{Einstein}, translating
into \emph{All locations in the Universe are equivalent; in particular, also the local, averaged density of stellar matter is to be
the same everywhere}. The second part of this statement is somewhat vague as the averaging process should refer to some physical scale.
The statement of the cosmological principle as expressed in \cite{Milne,Milne35} explicitly considers small-scale inhomogeneities and
is more relevant to the clumpy Universe.} To be more specific, we shall statistically quantify the approach to homogeneity in terms of
a decreasing clustering amplitude versus scale, as described by hierarchical clustering via the usual correlation functions \citep{Peeb80}.
If the galaxy distribution is given in real-distance space, then correlation functions indeed probe only the matter distribution inside
a given survey volume. However, observed galaxies are placed at their redshift coordinates which differ from the real distances by the
line-of-sight component of the peculiar velocity field $V=\vv\cdot\hvn$, giving rise to what is commonly described as redshift-space
distortions \cite{Jackson1972,k87,Fisher95b,Hamilton1998,Scoccimarro2004}. To first order in $\vv$, the continuity equation then implies
that the density field $\delta^{\rm s}$, which is inferred from the galaxy distribution in redshift space (hereafter {\sspa}),
includes the radial part of the divergence of $\vv$, i.e. $\bm{\grad}\cdot (V\hvn )$, in addition to the full divergence
$\bm{\grad}\cdot\vv$ which is proportional to the density field in real space. In linear theory of the standard gravitational paradigm
for structure formation, the velocity field is proportional to the gravitational force field which is generated not only by matter
fluctuations inside the survey volume, but also by external matter. Hence, unlike the full divergence which gives the local real-space
density, the radial divergence actually depends on the external mass distribution. A signature of this dependence should be imprinted
in the {\sspa} mass density as inferred from a redshift survey, allowing one to probe mass fluctuations on scales larger than the size
of the survey itself. Accessing these mass fluctuations on ``super-survey" scales thus offers an assessment of both the cosmological
principle and the fundamental physics of structure formation on these scales. 

In the absence of any preferred direction or location, the real-space density is a homogeneous and isotropic random field. However, the
corresponding {\sspa} galaxy distribution is anisotropic due to the additional line-of-sight displacement from distances to redshifts.
Therefore, traces of the external mass fluctuations would most easily be found in the apparent anisotropy of structure in {\sspa}. As
quantitative measures, we will consider moments of the {\sspa} correlation function, i.e.
$\xi^{\rm ss}(\vx,\vy) = \langle\delta^{\rm s}(\vx)\delta^{\rm s}(\vy)\rangle$, with respect to the angle between $\vy-\vx$
and $(\vx+\vy)/2$. Although we shall adopt standard gravity to describe the effect of external mass fluctuations, the general
approach presented below is applicable to alternative theories since in any such theory, the velocity field is likely a result of the
cumulative mass distribution as well. Hence, the effect can, in principle, also be regarded as a probe of the underlying gravitational
theory itself.

The outline of this paper is as follows: we begin with a description of the basic equations in section \ref{sec:equations}. In section
\ref{sec:fs}, we consider a full-sky case in which the observer is situated at the center of a finite-size sphere which is assumed to
represent the survey. Expanding the relevant fields in terms of spherical harmonics, which offer a convenient representation of the
velocity field generated by external mass fluctuations, we then present two examples which can be worked out analytically, and also 
explore the effect of external fluctuations on the {\sspa} density correlations within the standard $\Lambda$CDM cosmology. 
In section \ref{sec:do}, we will be concerned with the distant observer limit. In this case, we will assume a spherical survey
geometry of radius $R$ such that $R/D\ll 1$, where $D$ is the distance between the observer and the center of the survey. Note that
this configuration is more relevant to future, planned data products. Again, we will work with spherical harmonics (defined with
respect to the center of the survey region rather than the location of the observer), and further estimate the effect of external
fluctuations in this limit using a random realization of the density and velocity fields for the $\Lambda$CDM model. Finally, we
discuss possible implications for Fourier space analysis in section \ref{sec:imp}.

\section{The basic equations}
\label{sec:equations}
We shall adopt the following notation: the scale factor normalized to unity at the present epoch is $a(t)=1/(1+z)$, where $z$
is the redshift and $H(t)=\dot a/a$. The comoving coordinate and the physical peculiar velocity are denoted by $\vx$ and
$\vv=a\dd\vx /\dd t$, respectively. The mean density of the Universe is given by $\rho_{m}(t)$ and $\Omega_m=\rho_{m}/ \rho_c$, where
$\rho_{c}$ is the critical density. The real-space density contrast reads $\delta(t,\vx)=\rho/ \rho_{m}-1$ and its linear growing mode
is expressed as $\delta(t,\vx)=D(t)\delta(t_{0},\vx)$ \citep{Peeb80}. The growth factor of linear fluctuations is defined as
$f=\dd\log D/\dd\log a$, and the comoving distance to an object at redshift $z$ is given by
$x(z)=c\int_{t(z)}^{t_0}\dd t^{\prime}/a(t^{\prime})$. For simplicity, we restrict our analysis to the Newtonian equations of motion
for the evolution of large-scale structure \cite[e.g.,][]{Peeb80}, ignoring the general relativistic terms in the relation between the
density and velocity fluctuations \cite{chisari,wald12}. We also assume that linear theory is valid on the large scales considered below. 

The observed redshift $z_{\rm obs}$ of an object is related to its cosmological redshift $z$ through
\begin{equation}
z_{\rm obs} = z + (1+z)\frac{V}{c},
\end{equation}
where $V=\vv\cdot\hvn$ is the radial peculiar velocity. Note that we have neglected shifts due to the peculiar gravitational potential
and assumed that all redshifts are measured with respect to the rest frame of the cosmic microwave background (CMB) \cite{SW}. The
comoving distance estimated from $z_{\rm obs}$ is then
\begin{equation}
s = x(z_{\rm obs}) \approx x(z) + \frac{c}{H(z)}(z_{\rm obs} - z) = x + \frac{V}{aH}.
\end{equation} 
Therefore, we define the comoving redshift coordinate as 
\begin{equation}
\label{eq:xtos}
\vs = \vx + \frac{V}{aH} \hvn .
\end{equation}

Let $n^s(\vs)$ and $n(\vx)$ be the number densities of galaxies in {\sspa} and real space, respectively, where $\vs$ and $\vx$
are related by the mapping given in eq. \eqref{eq:xtos}. Introducing $\overline{n}$ as the mean number density of galaxies in the
survey (assumed to be the same in both spaces), we write $n^{s}(\vs) = \overline{n}\lbrack 1+\delta^{\rm s}(\vs)\rbrack$, and
accordingly, $n(\vx) = \overline{n}\lbrack 1+\delta(\vx)\rbrack$ for the density in real space. In the limit of $\delta \ll 1$ and
$\lvert V/(aHx)\rvert\ll 1$, substituting eq. \eqref{eq:xtos} into the corresponding continuity equation, i.e.
$n^s(\vs)\dd^3 s=n(\vx)\dd^3x$, yields
\begin{equation}
\label{eq:lins}
\delta^{\rm s} = \delta - \frac{1}{aH}\bm{\nabla}\cdot (V\hvn ),
\end{equation}
which links the density contrast $\delta$ to its counterpart $\delta^{\rm s}$ in {\sspa}. Note that the above expression also holds
for the corresponding mass density fields if one assumes that mass and galaxies share the same velocity field. The linearized Newtonian
equations of structure formation relate the mass density contrast $\delta_{m}$ to the divergence of the velocity field. For an unbiased
galaxy distribution, i.e. $\delta = \delta_{m}$, linear theory leads to the result
\begin{equation}
\label{eq:linx}
\delta = -\frac{1}{aHf}\bm{\nabla}\cdot\vv .
\end{equation}  
Considering a linear bias between $\delta$ and $\delta_m$, i.e. $\delta=b \delta_m$, the factor $f$ must be replaced by $\beta=f/b$. For
simplicity of notation, however, we will set $b=1$ for what follows and continue using $f$ rather than $\beta$.

The velocity field inside the observed survey volume, appearing in the real-space relation given by eq. \eqref{eq:linx}, may be decomposed
into the sum of two components, $\vv^{\rm I}$ and $\vv^{\rm X}$, which are generated by mass fluctuations inside and outside the survey,
respectively. The key point is now that inside the survey volume, we generally have $\bm{\nabla}\cdot V^{\rm X}\hvn\neq 0$ while  
$\bm{\nabla}\cdot\vv^{\rm X} = 0$. Therefore, $\delta^{\rm s}$ as measured from the survey contains information on the unobserved mass
distribution outside the sphere through the term
\begin{equation}
\delta^{\rm X} \equiv -\frac{1}{aH}\bm{\nabla}\cdot\left (V^{\rm X}\hvn\right ).
\end{equation}
In the following, we will assume a potential flow such that $\vv = aH(\bm{\nabla}\phi )$, i.e. the quantity
$\vu\equiv\bm{\nabla}\phi = \vv/(aH)$ denotes the \emph{comoving} peculiar velocity in distance units. In terms of $\phi$, the linear
velocity-density relation given by eq. \eqref{eq:lins} can be written as  
\begin{equation}
\label{eq:linz}
\frac{1}{f}\nabla^{2}\phi + \frac{1}{x^{2}}\frac{\partial}{\partial x}\left (x^{2}\frac{\partial \phi}{\partial x} \right ) = -\delta^{\rm s}.
\end{equation}
To first order, the spatial derivatives in eqs. \eqref{eq:lins} and \eqref{eq:linx} with respect to either $\vx$ or $\vs$ are equivalent,
and we also have $\delta^{\rm s}(\vs) = \delta^{\rm s}(\vx)$.

\section{The full-sky case}
\label{sec:fs}

Consider an all-sky survey limited to an outer radius $R$. The matter distribution outside the survey volume will generate a peculiar
velocity potential $\phi^{\rm X}$ which satisfies $\nabla^{2}\phi^{\rm X} = 0$ for $x < R$. However, the term describing the radial part
of the Laplacian in eq. \eqref{eq:linz} does generally not vanish for $\phi = \phi^{\rm X}$. To make further progress, we expand all
relevant fields in terms of spherical harmonics $Y_{lm}(\hvn )$, i.e. $\phi(\vx) = \sum_{l,m}\phi_{lm}(x) Y_{lm}(\hvn )$.
Introducing the expansion coefficients of the real-space density $\delta$ as
\begin{equation}
\delta_{lm}(x) = \int\dd\Omega\delta (x,\hvn )Y_{lm}(\hvn ),
\end{equation}
we obtain
\begin{equation}
\label{eq:phix}
\phi^{\rm X}_{lm}(x) = X_{lm}x^{l},
\end{equation}
where the constants $X_{lm}$ are given by \cite{BT}
\begin{equation}
\label{eq:Xlm}
X_{lm} = \frac{f}{2l+1}\int_{R}^{\infty}\frac{\dd a}{a^{l-1}}\delta_{lm}(a).
\end{equation}
Substituting this result back into eq. \eqref{eq:linz}, we therefore arrive at
\begin{equation}
\label{eq:dsx}
\delta^{\rm X}(\vx) = -\frac{1}{x^{2}}\frac{\partial}{\partial x}\left (x^{2}\frac{\partial\phi^{\rm X}}{\partial x} \right ) = -\sum_{l\geq 1}l(l+1)X_{lm}x^{l-2}Y_{lm}.
\end{equation}
The dipole term $(l=1)$, which formally diverges at $x=0$, corresponds to the bulk motion of the entire sphere as a result of the
gravitational pull caused by external density fluctuations. It can be eliminated by working with redshifts computed with respect to a
frame moving with this bulk flow (c.f. ref. \cite{ND94} regarding the Local Group versus CMB frames). Also, note that an external
quadruple moment $(l=2)$ generates a density which is constant with radius for $x<R$.

In preparation for the analysis conducted in section \ref{sec:lcdmall}, we will quantify the contribution of $\vv^{\rm X} $ to
$\delta^{\rm s}$ in terms of the correlation functions
\begin{equation}
\xi^{{\rm X}{\rm X}} = \left\langle\delta^{\rm X}(\vx)\delta^{\rm X}(\vy)\right\rangle ,
\qquad \xi^{\delta\rm X} = \frac{1}{2}\left (\xi^{\delta\rm X}_{12} + \xi^{\delta\rm X}_{21}\right ),
\end{equation}
where $\xi^{\delta\rm X}_{12} = \langle\delta(\vx)\delta^{\rm X}(\vy)\rangle$ and vice versa. The above functions emerge when
considering the correlation functions of $\delta^{\rm s}$ and $\tilde{\delta}^{\rm s} = \delta + \delta^{\rm X}$, leading to
$\langle\tilde{\delta}^{\rm s}\tilde{\delta}^{\rm s}\rangle = \langle\delta\delta\rangle + 2\xi^{\delta\rm X} + \xi^{\rm XX}$.
Note that the quantity $\tilde{\delta}^{\rm s}$ ignores distortions due to mass fluctuations within $R$, i.e. it explicitly neglects
the contribution coming from $V^{\rm I}$. Using eq. \eqref{eq:dsx}, it is then straightforward to show that
\begin{equation}
\xi^{{\rm X}{\rm X}}(\vx,\vy) = \sum_{l,m}Y_{lm}(\hvx )Y_{lm}^{*}(\hvy )l^{2}(l+1)^{2}(xy)^{l-2} X_{l}^{2}.
\end{equation}
Here $X_{l}^{2}\equiv \langle X_{lm}^{2}\rangle$ is the ensemble average of $X_{lm}^{2}$ which, as we will see shortly, is independent
of $m$. Squaring $X_{lm}$ as given by eq. \eqref{eq:Xlm} and averaging the resulting expression yields
\begin{equation}
\label{eq:cxx}
X_{l}^2 = \left(\frac{f}{2l+1}\right )^{2}\int_{R}^{\infty}\int_{R}^{\infty}\frac{\dd a\dd a^{\prime}}{a^{l-1}a^{\prime l-1}}\langle\delta_{lm}(a)\delta_{l^{\prime}m^{\prime}}(a^{\prime})\rangle .
\end{equation}
Since $\delta$ is an isotropic and homogeneous random field, its ensemble average $\langle\delta_{lm}(a)\delta_{l^{\prime}m^{\prime}}(a^{\prime})\rangle$
equals zero unless $l = l^{\prime}$ and $m = m^{\prime}$. Moreover, $\langle\delta_{lm}(a)\delta_{lm}(a^{\prime})\rangle$
is independent of $m$, which implies that this is also true for $\langle X_{lm}^2\rangle$. Introducing the Legendre polynomials
$\mathcal{P}_{l}$ and using the addition theorem, i.e.
\begin{equation}
\sum_{m}Y_{lm}(\hvx )Y_{lm}^{*}(\hvy ) = \frac{2l+1}{4\pi}\mathcal{P}_{l}(\mu ),\qquad \mu =\hvx\cdot\hvy ,
\end{equation}
we see that $\xi^{\rm XX}$ depends on $x$, $y$, and $\mu$ only. Thus we obtain
\begin{equation}
\label{eq:corrx}
\xi^{\rm XX}(x,y,\mu ) = \sum_{l\geq 1}\frac{2l+1}{4\pi}\mathcal{P}_{l}(\mu)C_{l}^{\rm XX},
\end{equation}
where $C_{l}^{\rm XX} = l^{2}(l+1)^{2}X^{2}_{l}(xy)^{l-2}$. Similar steps lead to
\begin{equation}
\xi^{\delta\rm X}_{12}(x,y,\mu ) = \sum_{l\geq 1}\frac{2l+1}{4\pi}\mathcal{P}_{l}(\mu)C_{l}^{\delta\rm X},
\end{equation}
where
\begin{equation}
\label{eq:corrmix}
C_{l}^{\delta\rm X} = -f\frac{l(l+1)}{2l+1}y^{l-2}\int_{R}^{\infty}\frac{\dd a}{a^{l-1}}\langle\delta_{lm}(x)\delta_{lm}(a)\rangle .
\end{equation}
In appendix \ref{sec:gen}, we will derive expressions relating $C^{\rm XX}$ and $C^{\delta\rm X}$ to the real-space density correlation
function $\xi = \langle\delta\delta\rangle$ and the power spectrum of density fluctuations.

\subsection{Two analytic examples}
\label{sec:analytic}

There are two particular functional forms of the underlying real-space correlation function, $\xi = \langle\delta\delta\rangle$, which
allow one to obtain closed analytic expressions of the corresponding $\xi^{\rm XX}$. These special cases will help us understand the
nature of $\xi^{\rm XX}$, its dependence on the size of the survey, $R$, and the underlying correlation $\xi$.

To begin with, consider the case $\xi = \sigma_{0}^{2}x_{0}/\lvert\vx - \vy\rvert$. A $\xi\propto 1/x$ dependence corresponds
to a power spectrum proportional to $k^{n}$, with a power index $n = -2$. Hence, on the large scales relevant to us, this form of $\xi$
is far from that of the concordance $\Lambda$CDM model \cite{wmap7}. Assuming that $l\geq 2$, the corresponding $X_{l}^{2}$ is then
given by (see appendix \ref{sec:cX})
\begin{equation}
\label{eq:clx}
X_{l}^{2} = \left (\frac{f}{2l+1}\right )^{2}\frac{4\pi\sigma_{0}^{2}x_{0}}{2l+1}\frac{2}{(2l-3)(2l-1)}\frac{1}{R^{2l-3}}.
\end{equation}
Substituting the above into eq. \eqref{eq:corrx}, we find
\begin{equation}
\xi^{\rm XX} = 2f^{2}\xi(R)\sum_{l}\left (\frac{l(l+1)}{2l+1} \right )^{2}\frac{\mathcal{P}_{l}(\hvx\cdot\hvy)}{(2l-3)(2l-1)}\left (\frac{xy}{R^{2}}\right )^{l-2}.
\end{equation}
Note that $\xi^{\rm XX} $ is not isotropic, i.e. it cannot be expressed as a function of the separation $\lvert\vx -\vy\rvert$
alone. Setting $\hvx\cdot\hvy = 1$ and $x = y = R/2$ the summation over $l\geq 2$ results in
$\xi^{\rm XX} = 1.08 f^{2}\xi(R)$.

The second form of $\xi$ allowing an analytic treatment is
$\xi(\lvert\vx -\vy\rvert) =\sigma_{0}^{2}x_{0}^{3}/\lvert\vx -\vy\rvert^{3}$. This choice corresponds to a
Harrison-Zel'dovich power spectrum $P(k)\propto k$, close to what is realized in the standard $\Lambda$CDM cosmology on large scales.
In this case, we find (again, see appendix \ref{sec:cX}) 
\begin{equation}
\langle\delta_{lm}(a)\delta_{lm}(a^{\prime})\rangle = {4\pi\sigma_{0}^{2}x_{0}^{3}}\frac{1}{a^{3}}\left (\frac{a^{\prime}}{a}\right )^{l} 
\frac{1}{1-(a^{\prime}/a)^{2}},
\end{equation}
which, when inserted into eq. \eqref{eq:cxx}, implies the presence of a logarithmic divergence in $X_{l}^{2}$ as a result of the
singularity at $a^{\prime} = a$. To make analytic progress, we simply approximate the last term in the above equation by an
appropriate finite geometric series. Within this approximation, we eventually get
\begin{equation}
\xi^{\rm XX} = 2f^{2}\xi(R)\sum_{l}S_{l}\frac{l^{2}(l+1)^{2}}{2l+1}\mathcal{P}_{l}(\hvx\cdot\hvy)\left (\frac{xy}{R^{2}}\right )^{l-2},
\end{equation}
where
\begin{equation}
S_{l} = \sum_{j=0}^{N}\frac{1}{(2l-1)(2l+2j+1)}
\end{equation}
is a general harmonic series evaluated up to some suitable choice of $N\gg 1$.

These two examples demonstrate the intuitive result that the correlation due to the external term is proportional to the mass correlation
function evaluated at a separation equal to the radius of the survey volume. 

\subsection{External contribution for an all-sky survey in the \boldmath{$\Lambda$}CDM model}
\label{sec:lcdmall}

To obtain a first rough estimate on how external fluctuations affect finite-size galaxy surveys, we turn to the standard $\Lambda$CDM
cosmology. Note that any deviations from the standard model on large scales may be gauged accordingly. Considering the analysis
below, we will restrict ourselves to the quadrupole and hexadecapole moments of the full {\sspa} correlation function given by
\begin{equation}
\xi_{2} = \frac{5}{2}\int_{-1}^{1}\xi^{\rm ss}(\vx,\vy)\mathcal{P}_{2}(\mu )\dd\mu,\qquad
\xi_{4} = \frac{9}{2}\int_{-1}^{1}\xi^{\rm ss}(\vx,\vy)\mathcal{P}_{4}(\mu )\dd\mu,
\end{equation}
respectively, where $\mu$ is now defined as the cosine of the angle between the vectors $\vy-\vx$ and
$(\vx+\vy)/2$, $\mathcal{P}_{2} = (3\mu^{2}-1)/2$, and $\mathcal{P}_{4} = (35\mu^{4}-30\mu^{2}+3)/8$. Similarly, we compute the
moments $\tilde{\xi}_{2}$ and $\tilde{\xi}_{4}$ of the correlation function
$\langle (\delta +\delta^{\rm X})(\delta +\delta^{\rm X})\rangle =\xi +2\delta^{\delta\rm X}+\xi^{\rm XX}$, i.e. the
correlation function of the previously introduced quantity $\tilde{\delta}^{\rm s}$ which excludes distortions due to mass
fluctuations inside radius $R$.

\begin{figure} 
\centering
\includegraphics[scale=0.45]{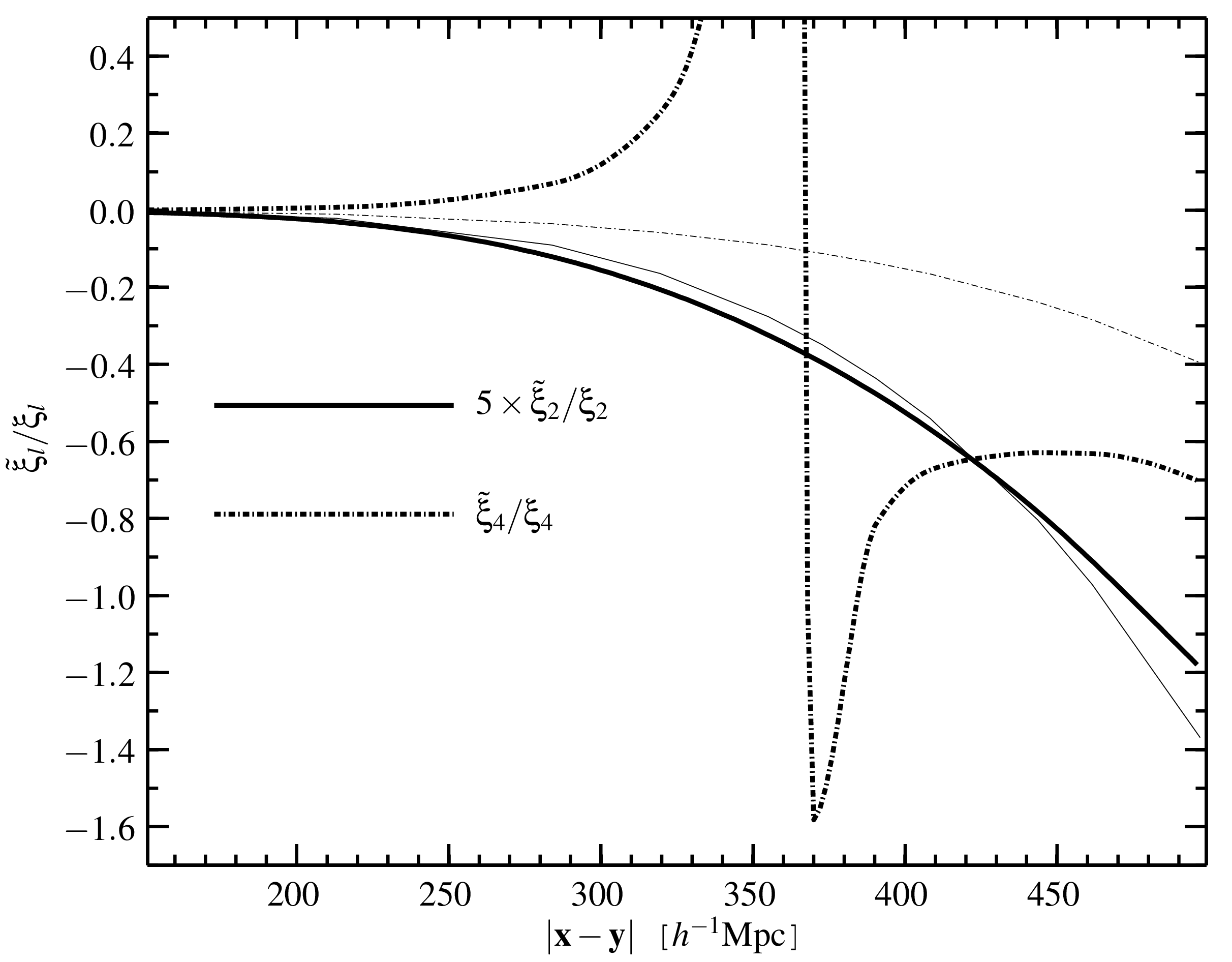}
\caption{The moment ratio of redshift space correlations computed with external contributions only, i.e. $V^{\rm X}$, and with the full
$V$ (see text for details). The solid and dashed lines correspond to the quadrupole and hexadecapole moments, respectively, plotted as
a function of separation between pairs of points with the same mean distance of approximately $350\hmpc$. The thick and thin lines
correspond to $f=1$ and $f=0.2$, respectively. For the sake of a better presentation, the quadrupole results have been multiplied by a
factor of $5$.}
\label{fig:Q}
\end{figure}

To quantify the effect of the external mass distribution, we compute the ratio $\tilde\xi_{l}/\xi_{l}$ for $l=2,4$ as a function of
separation. The ratios of the quadrupole and hexadecapole moments are a direct measure of the anisotropy induced by external
contributions, with a value of zero corresponding to a fully isotropic situation. Also, note that they are by far more sensitive
than the ratio of the monopole terms. Indeed, the ratio for the monopole turns out negligible, and we choose not to show it here.
For our calculations, we adopt a spatially flat cosmology with best-fit parameters based on the CMB anisotropies as measured by the
Wilkinson Microwave Anisotropy Probe (WMAP). Thus we assume a total mass density parameter $\Omega_m=0.266$, a baryonic density
parameter $\Omega_b=0.0449$, a value of $h=0.71$ for the Hubble constant in units of $100\kms {\rm Mpc}^{-1}$, a spectral index
$n_{s}=0.963$, and $\sigma_{8} = 0.80$ for the root mean square (rms) of linear density fluctuations within spheres of $8\hmpc$ radius
\cite{wmap7}. In addition, we work with a parametric form of the power spectrum which includes features due to baryonic acoustic
oscillations \cite{EH98}. The actual calculation of $\xi_2$ and $\xi_4$ is based on the expressions derived in appendix \ref{sec:gen},
setting $R=710\hmpc$ for the comoving limiting radius of the survey which corresponds to $z\approx 0.175$.

The results are presented in figure \ref{fig:Q} for correlations of pairs $\vx$ and $\vy$ with mean distance of
$\lvert\vx + \vy\rvert /2 = 356\hmpc$, corresponding to a redshift of $z\approx 0.086$. As can be easily seen, the figure
confirms the expectation that the impact of $V^{\rm X}$ becomes more significant as one moves to larger separations. The effect
of the external mass distribution is more pronounced in the hexadecapole ratio ($l=4$) which diverges at $\sim 356\hmpc$ due to
a change of sign in $\xi_{4}$ which is calculated from the full correlation function $\xi^{\rm ss}$. Moreover, we find that
$\lvert\tilde\xi_{4}/\xi_{4}\rvert$ remains larger than $0.5$ at separations $>400\hmpc$. Although the amplitude of the quadrupole
ratio is substantially smaller than that of the hexadecapole, it should be pointed out that the quadrupole is likely to be better
constrained by observations (see section \ref{sec:detect}).

\section{The distant observer limit}
\label{sec:do}
Now we will consider the distant observer limit (hereafter d.o.l.), i.e. a situation in which the survey volume is located at a
large distance from the observer such that the line of sight to any galaxy may be identified with a fixed direction chosen here
as the $x_{3}$-axis. To make the analytic calculations tractable, we additionally assume that the survey volume is given by a
sphere of radius $R$. The origin of the coordinate system is chosen to coincide with the center of the sphere, and the observer
is located at $x_{1}=x_{2}=0$ and $x_{3}=-\infty$. Thus $x=\lvert\vx\rvert$ denotes the distance of a point from the center of
the sphere. To begin with, we decompose the velocity field $\vu=\bm{\grad}\phi$ in terms of vector spherical harmonics which are
defined as $\vY_{lm}=Y_{lm}\hvn$ and $\vpsi_{lm}=x\bm{\nabla}Y_{lm}$ \citep{vsh}.\footnote{The following orthogonality conditions 
hold: $\int\dd\Omega\vY_{lm}\cdot\vY^{*}_{l^{\prime}m^{\prime}} = \delta^{K}_{ll^{\prime}}\delta^{K}_{mm^{\prime}}$,
$\int\dd\Omega\vpsi_{lm}\cdot\vpsi^{*}_{l^{\prime}m^{\prime}} = l(l+1)\delta^{K}_{ll^{\prime}}\delta^{K}_{mm^{\prime}}$, and
$\int\dd\Omega\vY_{lm}\cdot\vpsi^{*}_{l^{\prime}m^{\prime}} = 0$. Furthermore, $\vY_{lm}$ and $\vpsi_{lm}$ are orthogonal in the
usual three-dimensional sense, i.e. $\vY_{lm}\cdot\vpsi_{lm}=0$.} This yields
\begin{equation}
\label{eq:uvec}
\vu = \sum_{l,m}\left (\frac{\dd\phi_{lm}}{\dd x}\vY_{lm} + \frac{\phi_{lm}}{x}\vpsi_{lm}\right ),
\end{equation}
and the line-of-sight component takes the form
\begin{equation}
\label{eq:uz}
u_{3} = \hvx_{3}\cdot\vu =\sum_{l,m}\left (\mu\frac{\dd\phi_{lm}}{\dd x}Y_{lm} + {\phi_{lm}}\hvx_{3}\cdot\bm{\grad}Y_{lm}\right ),
\end{equation}
where we have introduced $\mu=x_3/x$. Substituting
\begin{equation}
\hvx_{3}\cdot\bm{\grad}Y_{lm} = \frac{\dd Y_{lm}}{\dd x_{3}} = \frac{1-\mu^{2}}{x}\frac{\dd Y_{lm}}{\dd\mu}
\end{equation}
into the above, we obtain
\begin{equation}
\label{eq:uzmu}
u_{3} = \sum_{l,m}\left\lbrack\mu\frac{\dd\phi_{lm}}{\dd x}Y_{lm} + \left (1-\mu^{2}\right )\frac{\phi_{lm}}{x}\frac{\dd Y_{lm}}{\dd\mu}\right\rbrack .
\end{equation}
For $\phi_{lm} = \phi^{\rm X}_{lm} = x^{l}X_{lm}$, where $X_{lm}$ is given by eq. \eqref{eq:phix}, the relation \eqref{eq:uzmu} reads 
\begin{equation}
u^{\rm X}_{3} = \sum_{l\geq 1,m}x^{l-1}X_{lm}\left\lbrack l\mu Y_{lm} + \left (1-\mu^{2}\right )\frac{\dd Y_{lm}}{\dd\mu}\right\rbrack .
\end{equation}
Using the definition of $Y_{lm}$ in terms of associated Legendre polynomials $\mathcal{P}^{m}_{l}$ and exploiting the recurrence relation
\begin{equation}
\left (\mu^{2}-1\right )\frac{\dd\mathcal{P}_{l}^{m}}{\dd\mu} = l\mu\mathcal{P}_{l}^{m}(\mu ) - (l+m)\mathcal{P}^{m}_{l-1}(\mu ),
\end{equation}
the last equation further simplifies to
 \begin{equation}
u^{\rm X}_{3} = \sum_{l\geq 0,m}X^{u}_{lm}x^{l}Y_{lm},
\end{equation}
with
\begin{equation}
X^{u}_{lm} = X_{l+1,m}\sqrt{\frac{2l+3}{2l+1}\left\lbrack (l+1)^{2} - m^{2}\right\rbrack}.
\end{equation}
Similarly, repeating the previous steps yields the induced {\sspa} density contrast $\delta^{\rm X}$ as
\begin{equation}
\label{eq:dolx}
\delta^{\rm X} = -\frac{\dd u^{\rm X}_{3}}{\dd x_{3}} = \sum_{l\geq 0,m}X^{d}_{lm}x^{l}Y_{lm},
\end{equation}
where
\begin{equation}
\label{eq:dolxx}
X^{d}_{lm} = -X_{l+2,m}\sqrt{\frac{2l+5}{2l+1}\left\lbrack (l+2)^{2} - m^{2}\right\rbrack\left\lbrack (l+1)^{2} - m^{2}\right\rbrack}. 
\end{equation}
The explicit dependence on $m$ is a direct consequence of the d.o.l. and the resulting fixed line-of-sight direction.  
The last two expressions imply that any multipole $l$ of $\delta^{\rm X}$ is affected by the $l+2$ component of the external
mass distribution. For example, the quadrupole of the external mass contributes to the monopole of $\delta^{\rm X}$. In the
full-sky case with an observer placed at the center of a spherical survey volume, multipoles of the external mass distribution
contribute to the internal {\sspa} density distortions at the same level, i.e. there is no ``mixing". This behavior can be
understood from the different additional density terms associated with the external mass distribution which are 
$\partial (x^{2}\partial\phi)/x^{2}\partial x$ for the latter case, and $\dd^{2}\phi/\dd x_{3}^{2}$ in the d.o.l. approximation.

\subsection{External mass contribution for the \boldmath{$\Lambda$}CDM model in the d.o.l.}
\label{sec:mc}

As we have seen in the last section, the expressions for $\delta^{\rm X}$ in the d.o.l., i.e. eqs. \eqref{eq:dolx} and \eqref{eq:dolxx},
involve an explicit dependence on $m$, which greatly complicates the calculation of the relevant correlation functions and their moments.
We therefore resort to a calculation of the moments based on a numerical realization of the random density field, $\delta$, together
with the corresponding fields $\vv$ and $\delta^{\rm s}$. Adopting the same $\Lambda$CDM power spectrum as in section \ref{sec:lcdmall},
we generated a linear Gaussian realization of the real-space density field on a cubic grid with $1024^{3}$ grid points, sampling a total
volume of $(4000\hmpc)^{3}$. The corresponding velocity field inside the whole box was obtained using linear theory with $f=1$ and 
assuming periodic boundary conditions. This velocity field is used to recover the {\sspa} density $\delta^{\rm s}=\delta -\dd u_{3}/\dd x_{3}$
inside the box, where as before, the line-of-sight direction is assumed as parallel to the $x_{3}$-axis.  
The whole box is then divided into $8$ sub-boxes, with each sub-box covering $2000\hmpc$ on the side. At each grid point in any of the
sub-boxes, the real-space density is taken from the corresponding point in the whole box. However, we now consider two estimates of the
{\sspa} density. The first is simply $\delta^{\rm s}$ computed from the whole box as described above. The second estimate is obtained as
follows: for each sub-box, we compute the velocity field assuming zero padding of the sub-box inside the whole box, but still imposing
periodic boundary conditions on the whole box. This velocity field is used to compute the {\sspa} density $\delta_{\rm zp}^{\rm s}$ inside
the sub-box. After this step, we proceed with computing the correlation functions in the sub-boxes for the {\sspa} densities obtained with
and without zero padding. The correlation functions are obtained by directly averaging over pairs in each sub-box, as a function of the
line-of-sight and projected separations. Finally, the quadrupole and hexadecapole moments are found by means of direct numerical
integration over the angle between the separation vector and the line-of-sight direction.

\begin{figure} 
\centering
\includegraphics[scale=0.46]{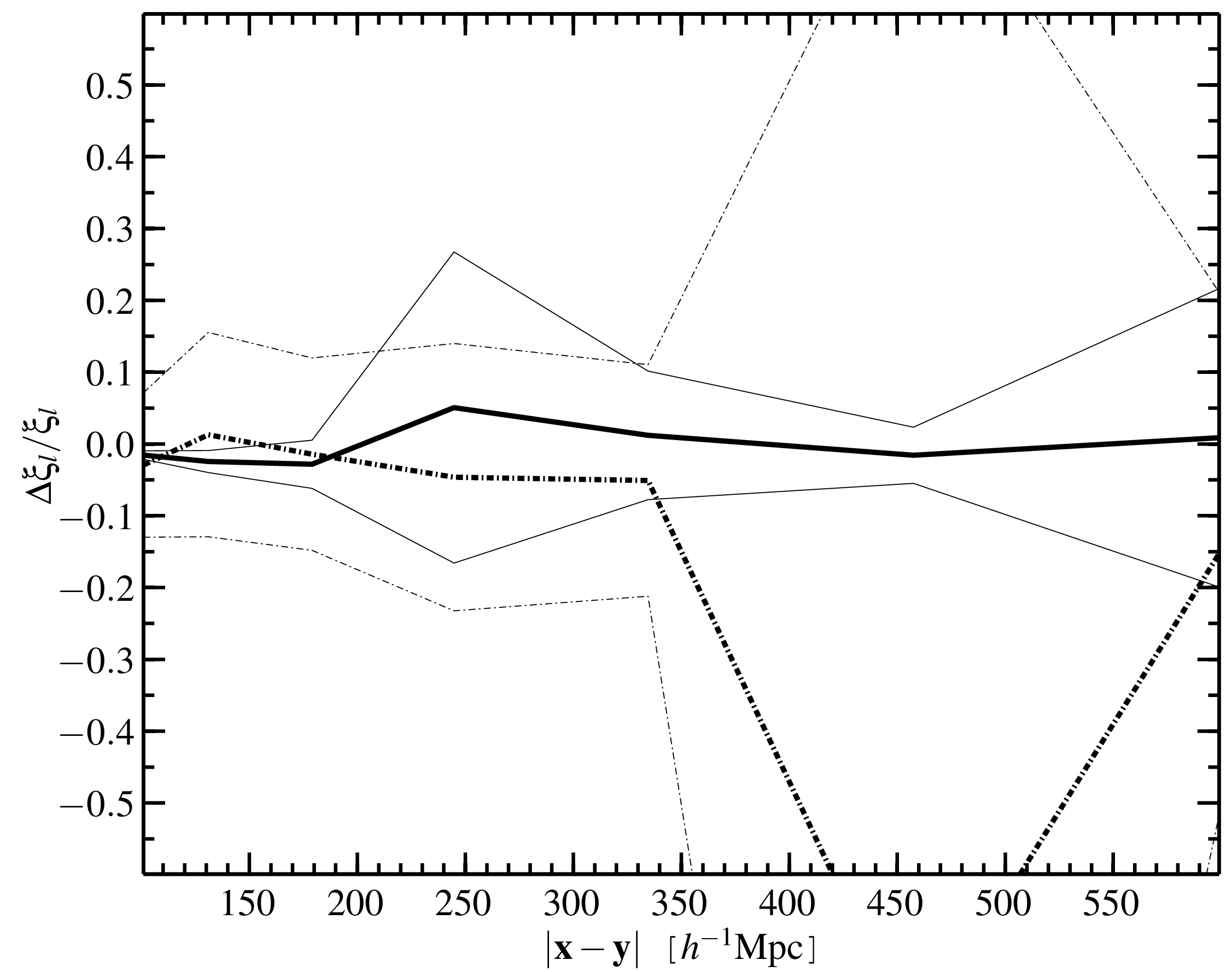}
\caption{Resulting relative differences of multipole moments: the quadrupole and hexadecapole moments are shown as solid and dash-dotted
lines, respectively. Thick lines correspond to the mean of the ratios computed from the 8 sub-boxes while the thin lines indicate the
$\pm\sigma$ rms scatter which is also obtained from the 8 sub-boxes.}
\label{fig:Qdol}
\end{figure}
 
Figure \ref{fig:Qdol} shows the relative difference of the moments computed with and without the contribution of external mass fluctuations. 
It is important to note that the ratios $\Delta\xi_{l}/\xi_{l}$ in this figure have a different meaning from the results plotted in figure
\ref{fig:Q}. For technical reasons, the calculated moment ratios in figure \ref{fig:Q} are based on the full $\xi^{\rm ss}$ (including
external and internal contributions) and the correlation function of $\tilde{\delta}^{\rm s}$ which has zero contribution from the velocity
field generated by the internal mass density. Moreover, figure \ref{fig:Q} only refers to pairs of points with a mean distance of $356\hmpc$
while all pairs are included in the calculation of moments used in figure \ref{fig:Qdol}.

Instead of illustrating the relative differences
for each sub-box individually, figure \ref{fig:Qdol} shows the mean and $1\sigma$ rms scatter of the ratios obtained from the 8 sub-boxes.
Again, we find that the effect is more pronounced in the hexadecapole as compared to the quadrupole moment. Since the volume of each sub-box
is comparable to the effective volume of planned and ongoing redshift surveys, the resulting scatter of a few percent and more should
appropriately reflect the expected additional contribution due to the externally sourced field $\delta^{\rm X}$ in a $\Lambda$CDM cosmology.

\subsection{Detecting external mass contributions in next-generation surveys}
\label{sec:detect}
\begin{figure} 
\centering
\includegraphics[scale=0.62]{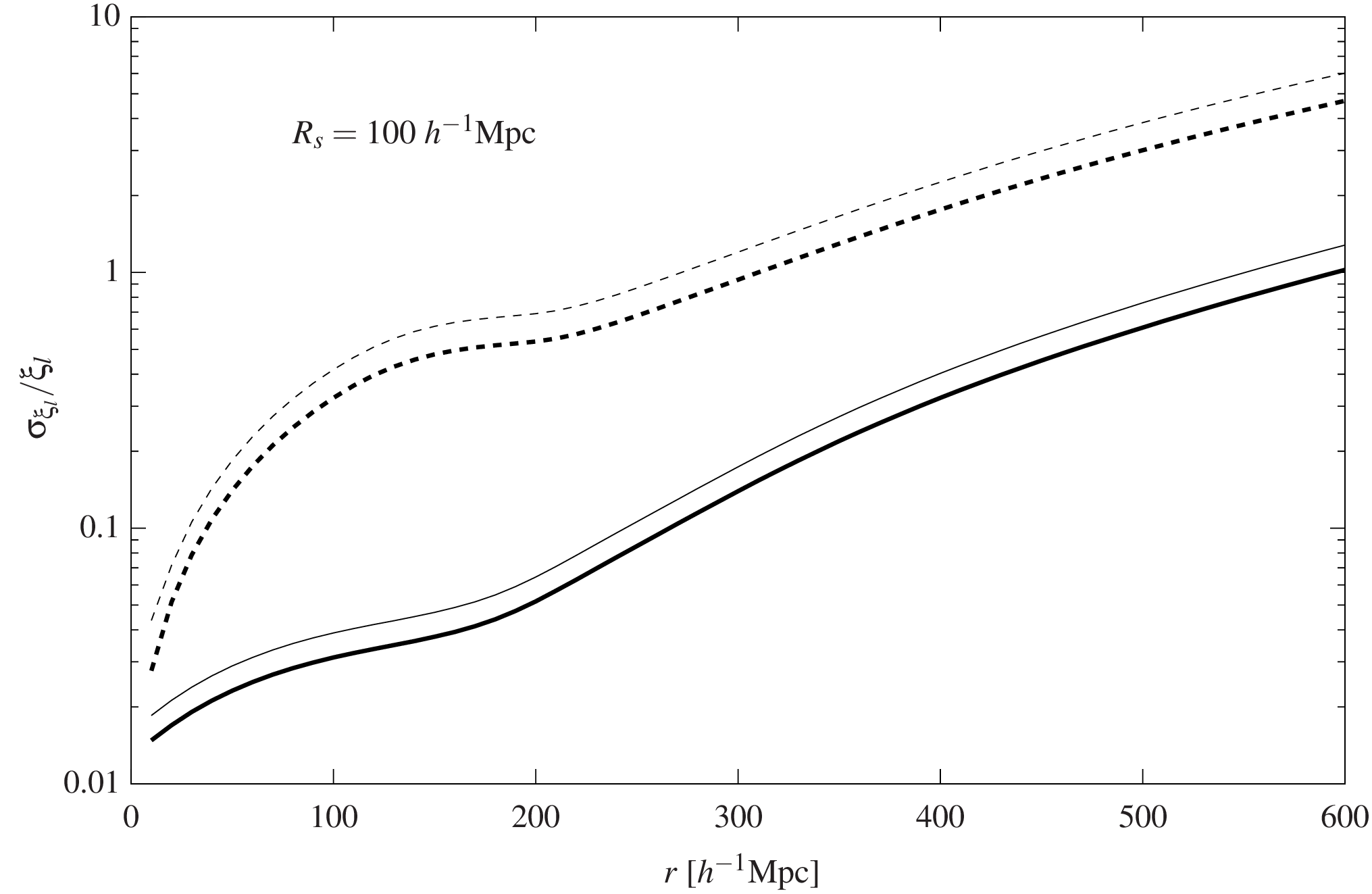}
\caption{Expected noise-to-signal ratios $\sigma_{\xi_{l}}/\xi_{l}$ for the two bins described in the text: assuming a density
smoothing scale of $R_{s}=100h^{-1}$Mpc, the ratios for the quadrupole and hexadecapole are shown as solid and dashed lines,
respectively. The adopted redshift bins at $\overline{z}_{1}=0.89$ (for BigBOSS) and $\overline{z}_{2}=1.07$ (for Euclid)
correspond to thin and thick lines, respectively.}
\label{fig:NSdol}
\end{figure}

In this section, we will address whether the contribution due to external mass fluctuations will become detectable by directly measuring the
quadrupole and hexadecapole moments in future galaxy redshift surveys. Using the d.o.l. approximation, we start by writing the multipole
moments of the {\sspa} correlation function as \cite{Hamilton1992}
\begin{equation}
\begin{split}
\xi_{2}\equiv\xi^{\rm ss}_{2}(r) &= -\left (\frac{4f}{3} + \frac{4f^{2}}{7}\right )\int\frac{\dd^{3}k}{(2\pi )^{3}}P(k)j_{2}(kr),\\
\xi_{4}\equiv\xi^{\rm ss}_{4}(r) &= \frac{8f^{2}}{35}\int\frac{\dd^{3}k}{(2\pi )^{3}}P(k)j_{4}(kr),
\end{split}
\label{eq:detect1}
\end{equation}
where we have again set $b=1$ (see section \ref{sec:equations}). Considering a finite survey volume $V$, multipole moments of $\xi^{\rm ss}$
are typically computed from previous estimates of the full correlation function. The noise $\sigma_{\xi_{l}}$ associated with a measurement of
$\xi_{l}$ can be calculated in terms of the covariance between the corresponding power spectrum multipoles, i.e.
\begin{equation}
\sigma^{2}_{\xi_{l}}(r) = \int\frac{\dd^{3}k}{(2\pi )^{3}}\int\frac{\dd^{3}k^{\prime}}{(2\pi )^{3}}\left\langle\Delta
\hat{P}^{\rm ss}_{l}(\vk )\Delta\hat{P}^{\rm ss}_{l}(\vk^{\prime})\right\rangle j_{l}(kr)j_{l}(k^{\prime}r),
\label{eq:detect2}
\end{equation}
Here $\Delta\hat{P}^{\rm ss}_{l}(\vk )\Delta\hat{P}^{\rm ss}_{l}(\vk^{\prime})=(\hat{P}^{\rm ss}_{l}(\vk )-\langle\hat{P}^{\rm ss}_{l}
(\vk )\rangle )(\hat{P}^{\rm ss}_{l}(\vk^{\prime})-\langle\hat{P}^{\rm ss}_{l}(\vk^{\prime})\rangle )$, the estimator of the multipole
$P^{\rm ss}_{l}$ is $\hat{P}^{\rm ss}_{l}(\vk )$ (see appendix \ref{app:var}), and $j_{l}$ are the usual spherical Bessel function of
the first kind. Accounting for shot noise and assuming that fluctuations of the density field on the large scales we are concerned with
here are governed by Gaussian statistics, the above expression takes the form (again, see appendix \ref{app:var})
\begin{equation}
\sigma^{2}_{\xi_{l}}(r) = \frac{(2l+1)^{2}}{V}\int\frac{\dd^{3}k}{(2\pi )^{3}}\left\lbrack\mathcal{C}_{l4}P^{2}(k) + 2\mathcal{C}_{l2}P(k)
\overline{n}^{-1} + \mathcal{C}_{l0}\overline{n}^{-2}\right\rbrack j^{2}_{l}(kr),
\label{eq:detect3}
\end{equation}
where we have defined
\begin{equation}
\mathcal{C}_{ll^{\prime}} = \int_{-1}^{1}\dd\mu_{\vk}\mathcal{P}_{l}^{2}\left (\mu_{\vk}\right )\left (1+f\mu_{\vk}^{2}\right )^{l^{\prime}}.
\label{eq:detect3b}
\end{equation}
As examples for future large (spectroscopic) redshift surveys, let us consider the planned BigBOSS experiment \cite{bigboss2011} and the
Euclid spectroscopic survey \cite{euclid09}. Based on the ground, the first one will observe around $20$ million galaxies up to $z\sim 1.7$
over more than $30$\% of the sky. Covering a similar fraction of the sky, the number of Euclid galaxies with redshifts $0.7<z<2.0$ is even
expected to be as high as $\sim 10^{8}$. For simplicity, we choose single redshift bins with $0.5<z_{1}<1.3$ and $0.7<z_{2}<1.6$ for
BigBOSS and Euclid, respectively. Figure \ref{fig:NSdol} shows the expected $\sigma_{\xi_{l}}/\xi_{l}$ ratios for both the quadrupole
(solid lines) and the hexadecapole (dashed lines). Note that the results assume a density smoothing scale of $R_{s} = 100h^{-1}$Mpc to
reduce the noise contribution from smaller scales which are not relevant to the signal from external fluctuations. Concerning the actual
calculation, we have adopted mean bin redshifts (number densities) of $\overline{z}_{1}=0.89$
($\overline{n}_{1}=5.89\times 10^{-4}h^{3}$Mpc$^{-3}$) and $\overline{z}_{2}=1.07$ ($\overline{n}_{2}=1.44\times 10^{-3}h^{3}$Mpc$^{-3}$)
which are based on the expected reference distributions given in \cite{bigboss2011} and \cite{Amendola2013}. For computing the power
spectrum, we have again used the $\Lambda$CDM cosmology introduced in section \ref{sec:lcdmall}. Looking at the results of figure
\ref{fig:NSdol}, we see that the quadrupole at $200$--$250h^{-1}$Mpc should be constrained up to $6$--$10$\% and $5$--$8$\% in the two
different bins. Comparing this to figure \ref{fig:Qdol}, there is a good chance of detecting signatures of external fluctuations on this
scale which could amount to relative deviations of up to roughly $10$--$20$\%. In particular, this should be true if one considers the
ratio to the monopole, which further decreases the errors by alleviating the dominant contribution of cosmic variance. As for the
hexadecapole moment, the situation is much worse. The noise level at separations $r>150h^{-1}$Mpc already makes up more than $50$\% and
reaches $90$--$100$\% at $r\sim 300h^{-1}$Mpc. Unless the external fluctuations cause deviations significantly above the $100$\% level,
this makes their detection through the hexadecapole very hard to virtually impossible. Again, looking at ratios of multipole moments
might change the prospects of detecting a corresponding signal for the better.

While the above is complete at the linear level, nonlinear clustering introduces uncorrelated virial motion which may not be negligible,
even at very large separations \cite{Scoccimarro2004}. This gives rise to a scale- and angle-dependent suppression of the {\sspa} power
spectrum relative to the Kaiser approximation \cite{k87}. Consequently, there will be corrections to the multipole moments $\xi_{l}$
calculated from eq. \eqref{eq:detect1} which might be large enough to hide any effects due to external fluctuations. A brief calculation
shows that these nonlinear corrections do not pose a problem for our previous findings on $\xi_{2}$ since they remain smaller than
$0.2$--$0.3$\% at separations $\gtrsim 200h^{-1}$Mpc. As for the hexadecapole $\xi_{4}$, the relative differences are probably much larger
on these scales, in which case the apparent remedy is to include nonlinear contributions as accurate as possible when modeling redshift
space distortions. 
Similarly, other effects which have an impact on the {\sspa} correlation function such as nonlinear (or scale-dependent) galaxy biasing
and magnification bias \cite{Hui2007, Hui2008} should be taken into account within a full clustering analysis. Generated by gravitational
lensing effects along the line-of-sight, the latter interestingly also leads to additional anisotropies in the observed correlation
function. As opposed to the contribution of external fluctuations, characterized by a specific dependence on the distance from the survey
center, the changes due to magnification bias are strongest in the line-of-sight direction, possibly allowing one to separate the two
effects in a given data set.

\section{Possible implications for power spectrum analysis} 
\label{sec:imp}
In Fourier space, the linear real-space relation between the density, $\delta$, and the velocity potential, $\phi$, is given by
$k^{2}\phi_{\vk}=f\delta_{\vk}$. In the d.o.l., this relation gives
$\delta^{T}_{\vk}=k_{3}^{2}\phi_{\vk} =f({\vk}\cdot\hvx_{3})^{2}\delta_{\vk}$ as the Fourier space equivalent of
$\delta^{T} = -\dd u_{3}/\dd x_{3}$. Therefore, it follows that
\begin{equation}
\label{eq:dvk}
\delta^{\rm s}_{\vk} = \delta_{\vk}(1+f\mu_{\vk}^{2}), 
\end{equation}
where $\mu_{\vk} = \hvk\cdot\hvx_{3}$ and we have ignored corrections due to small-scale incoherent motions. Usually,
$\delta^{\rm s}_{\vk}$ is estimated by applying the Fast Fourier Transform (FFT) on data points in a box which is embedded
in the survey region. Since the FFT employs periodic boundary conditions, eq. \eqref{eq:dvk} turns into 
\begin{equation}
\label{eq:dvkx}
\delta^{\rm s}_{\vk} = \delta_{\vk}(1+f\mu_{\vk}^{2}) +\mu_{\vk}^{2}k^{2}\phi^{\rm X}_{\vk} 
\end{equation}
for Fourier modes which are computed with the FFT. Here $\phi^{\rm X}_{\vk}$ is the Fourier transform of the velocity potential
induced by mass fluctuations \emph{external} to the box. The other term on the right-hand side of eq. \eqref{eq:dvkx} represents
modes inside the box which in addition to $\delta_{\vk}$, also comprise the distortion term $f\mu_{\vk}^{2}\delta_{\vk}$. It is
crucial to note that $\phi^{\rm X}(\vx)$ can be captured extremely well by the FFT inside the box. After all, this transform is
just a linear mapping between two sets of numbers. Since the FFT of $\nabla^{2}\phi^{\rm X}$ is $-k^{2}\phi^{\rm X}_{\vk}$ does
obviously not vanish, however, we conclude that $\phi^{\rm X}_{\vk}$ simply cannot yield $\nabla^{2}\phi^{\rm X}=0$. This is a
subtle point regarding the use of FFTs, and we encourage the reader to consult the mathematical literature for additional insight.
Therefore, using eq. \eqref{eq:dvkx} to infer $f$ from redshift distortions without the term involving $\phi^{\rm X}_{\vk}$ is
strictly incorrect. As we have seen from the previous sections, the term $\delta^{\rm X}$ could change the quadrupole and hexadecapole
moments in the $\Lambda$CDM model by a few per cents. Since the monopole remains basically unaffected, this should also hold true
for the corresponding multipole ratios. Although we do not provide a rigorous calculation, we thus expect that neglecting the
term $\phi^{\rm X}_{\vk}$ could change the estimate of $f$ also by a few per cents or so. Clearly, future work should address this
particular question in more detail.

\section{Discussion}

Among other successes, the standard $\Lambda$CDM cosmology has proven remarkably consistent with large-scale observations of
the galaxy distribution and the temperature fluctuations of the CMB \cite[e.g.][]{planck13,bossbao}. Nonetheless, anomalies in
the temperature spectrum persist \citep[e.g][]{planckanomalies}, which may or may not indicate the need for modifications of
the basic paradigm. Measured anisotropies in the CMB temperature map originate from fluctuations at the last scattering surface
at high redshift and cumulative effects along the traveling path of photons. Despite their impressive precision measurements,
however, CMB anisotropies may still miss important information regarding density fluctuations on scales larger than those
dictated by the size of redshift surveys.

In this work, we argue that such super-survey scales may actually be probed by the redshift space density obtained from the
distribution of galaxies in the redshift surveys. This information could, in principle, be used as a probe of both fluctuations
on super-survey scales and the fundamental theory of structure formation on large scales. Within the standard paradigm of
structure formation, external fluctuations induce a velocity potential of the form $x^{l}Y_{lm}(\hvn)$ inside a given survey
volume and thus lead to additional anisotropies in the observed correlations. The specific dependence on the distance $x$ from
the survey center could be used to extract information by either directly fitting these functional forms to the redshift space
density field or by measuring the correlation functions at various depths. 
 
Standard methods for estimating the growth rate from redshift space distortions are based on theoretical relations and numerical
results which either assume a survey of infinite size or a periodic box where the density fluctuations responsible for the
velocity field are all contained within the survey volume. The effect of a velocity field generated by external fluctuations is
not explicitly included in these relations, which could lead to a non-negligible uncertainty in the estimates of the growth rate. 

The analysis presented here is mostly relevant for future large galaxy surveys such as BigBOSS \cite{bigboss2011} and the Euclid
mission \cite{euclid09} in which the correlation functions on large scales can be estimated to very high accuracy due to the large
number of available galaxies. 

\acknowledgments
This research was supported by the I-CORE Program of the Planning and Budgeting Committee, THE ISRAEL SCIENCE FOUNDATION (grants No. 1829/12
and No. 203/09), the German-Israeli Foundation for Research and Development, and the Asher Space Research Institute. M.F. is supported by a
fellowship from the Minerva Foundation.

\appendix
\section{Solution of the linear velocity-density relation in terms of spherical harmonics}
For distinguishing between internal and external contributions to $\phi$, it turns out suitable to express the angular
dependence of $\phi(x,\hvn)$ in terms of spherical harmonics $Y_{lm}$. We therefore expand 
\begin{equation}
\phi(\vx) = \sum_{l,m}\phi_{lm}(x)Y_{lm}(\hvn ),
\end{equation}
and write the well-known solution to Poisson's equation $\nabla^{2}\phi =-f\delta$ as \cite{BT}
\begin{equation}
\phi_{lm}(x) = -\frac{f}{2l+1}\left (\frac{1}{x^{l+1}}\int_{0}^{x}\dd a a^{l+2}\delta_{lm}(a) + x^{l}\int_{x}^{R}\frac{\dd a}{a^{l-1}}\delta_{lm}(a)\right ) - x^{l}X_{lm},
\end{equation}
where 
\begin{equation}
X_{lm} = \frac{f}{2l+1}\int_{R}^{\infty}\frac{\dd a}{a^{l-1}}\delta_{lm}(a) 
\end{equation}
is the explicit contribution from mass fluctuations outside the considered volume, i.e. $x>R$.

\section{Computing redshift space correlations for generic power spectra}
\label{sec:gen}
We start from \citep{Fisher95b},
\begin{equation}
\label{eq:fish}
\begin{split}
\xi\left (\lvert\vx-\vy\rvert\right ) &= \frac{1}{(2\pi)^{3}}\int\dd^3 kP(k){\rm e}^{-i\vk\cdot (\vx-\vy)}\\
&= \frac{2}{\pi}\sum_{l,m}Y_{lm}(\hvx)Y^{*}_{lm}(\hvy)\int_{0}^{\infty}\dd k k^{2}P(k)j_{l}(kx)j_{l}(ky)\\
&= \frac{1}{2\pi^{2}}\sum_{l}(2l+1)\mathcal{P}_{l}(\hvx\cdot\hvy)\int_{0}^{\infty}\dd k k^{2}P(k)j_{l}(kx)j_{l}(ky),
\end{split}
\end{equation}
where we have used $\sum_{m}Y_{lm}(\hvx)Y^{*}_{lm}(\hvy) = (2l+1)\mathcal{P}_{l}(\hvx\cdot\hvy)/4\pi$ and
\begin{equation}
{\rm e}^{i\vk\cdot\vx} = 4\pi\sum_{l,m}i^{l}j_{l}(kx)Y_{lm}(\hvx)Y_{lm}^{*}\left ({\hat{\vk}}\right ).
\end{equation}
To evaluate the term $X_{l}^{2}$ in eq. \eqref{eq:cxx}, we further need
\begin{equation}
\label{eq:dlm}
\begin{split}
\left\langle\delta_{lm}(a)\delta_{lm}(a^{\prime})\right\rangle &= \left\langle\int\dd\Omega\dd\Omega^{\prime}\delta(\va )
\delta(\va^{\prime})Y_{lm}(\hva )Y^{*}_{lm}(\hva^{\prime})\right\rangle\\
& = \int\dd\Omega\dd\Omega^{\prime}\xi (\lvert\va -\va^{\prime}\rvert )Y_{lm}(\hva )Y^{*}_{lm}(\hva^{\prime})\\
& = 2\pi\int\dd\mu^{\prime}\xi (\lvert\va -\va^{\prime}\rvert )\mathcal{P}_{l}(\mu^{\prime}),
\end{split}
\end{equation}
where the last step results from averaging the product of $Y_{lm}$'s over all $2l+1$ $m$-dependent terms to get
$\mathcal{P}_{l}(\mu^{\prime})/4\pi$ with $\mu^{\prime} = \hat{\va}\cdot\hat{\va}^{\prime}$. Substituting eq. \eqref{eq:fish} into
the above and using the orthogonality relation
$\int\dd\mu^{\prime}\mathcal{P}_{l}\mathcal{P}_{l^{\prime}} = 2\delta^{K}_{ll^{\prime}}/(2l+1)$, one obtains
\begin{equation}
\label{eq:dlmk}
\begin{split}
\left\langle\delta_{lm}(a)\delta_{lm}(a^{\prime})\right\rangle &= \frac{1}{\pi}\int_{-1}^{1}\dd\mu^{\prime}\sum_{l^{\prime}}
(2l^{\prime}+1)\mathcal{P}_{l}(\mu^{\prime})\mathcal{P}_{l^{\prime}}(\mu^{\prime})\int_{0}^{\infty}\dd k k^{2}P(k)j_{l^{\prime}}(k a)j_{l^{\prime}}(k a^{\prime})\\
&= \frac{2}{\pi}\int_{0}^{\infty}\dd k k^2P(k) j_{l}(k a) j_{l}(k a^{\prime}).
\end{split}
\end{equation}
Hence, we arrive at
\begin{equation}
\begin{split}
X_{l}^{2} &= \left (\frac{f}{2l+1}\right )^{2}\int_{R}^{\infty}\int_{R}^{\infty}
\frac{\dd a\dd a^{\prime}}{a^{l-1}a^{\prime l-1}}\langle\delta_{lm}(a)\delta_{lm}(a^{\prime})\rangle\\
&= \frac{2}{\pi}\left (\frac{f}{2l+1}\right )^{2}R^{-2(l-1)}\int_{0}^{\infty}\dd k k^{2}P(k)k^{-2}\lbrack j_{l-1}(kR)\rbrack^{2},
\end{split}
\end{equation}
where we have exploited that
\begin{equation}
F(w)=\int_{w}^{\infty}\dd z z^{1-l}j_{l}(z) = w^{1-l}j_{l-1}(w).
\end{equation}
Similarly, one finds
\begin{equation}
\int_{R}^{\infty}\frac{\dd a}{a^{l-1}}\langle\delta_{lm}(x)\delta_{lm}(a^{\prime})\rangle =
\frac{2}{\pi}R^{1-l}\int_{0}^{\infty}\dd k k^2 P(k) k^{-1}j_{l}(kx)j_{l-1}(kR),
\end{equation}
which may be used to evaluate the expression for $C_{l}^{\delta\rm X}$ given by eq. \eqref{eq:corrmix}.

In the following, we wish to estimate the amplitude of $\langle\delta^{\rm s}\delta^{\rm s}\rangle$, i.e. the {\sspa}
correlation function, including the contribution from both internal and external fields. Let
\begin{equation}
\delta^{T}\equiv\delta^{\rm s} - \delta = -\grad^{2}\phi\vert_{\rm radial} =
-\frac{1}{x^{2}}\frac{\partial}{\partial x}\left (x^{2}\frac{\partial}{\partial x}\phi\right )
\end{equation}
be the full contribution to {\sspa} distortions from the entire space. Expanding $-f \delta=\grad^2\phi$
in terms of $Y_{lm}$ yields
\begin{equation}
-f\delta_{lm} = \frac{1}{x^{2}}\frac{\partial}{\partial x}\left (x^{2}\frac{\partial\phi_{lm}}{\partial x}\right ) -l(l+1)\frac{\phi_{lm}}{x^{2}},
\end{equation}
and thus we obtain
\begin{equation}
\delta^{T}_{lm} = f\delta_{lm} -l(l+1)\frac{\phi_{lm}}{x^{2}}.
\end{equation}
Using that $\delta^{\rm s}=\delta +\delta^{T}$, we now seek an expression for
$\xi^{\rm ss}(\vx,\vy) = \langle\delta^{\rm s}(\vx)\delta^{\rm s}(\vy)\rangle$. To this end, we write
\begin{equation}
\xi^{\rm ss}(\vx,\vy) = \frac{1}{4\pi}\sum_{l}(2l+1)\mathcal{P}_{l}(\hat{\vx}\cdot\hat{\vy})C_l^{\delta^{\rm s} \delta^{\rm s}}(x,y),
\end{equation}
where 
\begin{equation}
\begin{split}
C_{l}^{\delta^{\rm s}\delta^{\rm s}}(x,y) &= \langle\delta^{\rm s}_{lm}(x)\delta_{lm}^s(y)\rangle\\
&= (1+f)^{2}\langle\delta_{lm}(x)\delta_{lm}(y)\rangle -(1+f)l(l+1)\left (\frac{1}{x^{2}}+\frac{1}{y^{2}}\right )\\
&\quad\times\langle\phi_{lm}(x)\delta_{lm}(y)\rangle + \frac{l^{2}(l+1)^{2}}{x^{2}y^{2}}\langle\phi_{lm}(x)\phi_{lm}(y)\rangle .
\end{split}
\end{equation}
Considering this result, one should note that $\phi\propto f$ if expressed as a solution to $-f \delta=\grad^2\phi$. The
first term, i.e. $\langle\delta_{lm}(x)\delta_{lm}(y)\rangle$, is given by eq. \eqref{eq:dlmk}, and similar steps
as above eventually lead to
\begin{align}
\langle\delta_{lm}(x)\phi_{lm}(y)\rangle &= \frac{2f}{\pi}\int_{0}^{\infty}\dd k P(k)j_{l}(kx)j_{l}(ky),\\
\langle\phi_{lm}(x)\phi_{lm}(y)\rangle &= \frac{2f^{2}}{\pi}\int_{0}^{\infty}\dd k k^{-2}P(k)j_{l}(kx)j_{l}(ky), 
\end{align}
where we have used $\langle\phi_{\vk}\delta_{\vk}\rangle =f\langle\lvert\delta_{\vk}\rvert^{2}\rangle/k^{2}$ which
results from $f\delta_{\vk} = k^{2}\phi_{\vk}$. Finally, expanding the autocorrelation function
$\xi^{TT}(\vx,\vy) \equiv \langle\delta^{T}(\vx)\delta^{T}(\vy)\rangle$ as
\begin{equation}
\xi^{TT}(\vx,\vy) = \frac{1}{4\pi}\sum_{l}(2l+1)\mathcal{P}_{l}(\hat{\vx}\cdot\hat{\vy})C_{l}^{\delta^{T}\delta^{T}}(x,y)
\end{equation}
yields 
\begin{equation}
\begin{split}
C^{\delta^{T}\delta^{T}}_{l}(x,y) &= f^{2}\langle\delta_{lm}(x)\delta_{lm}(y)\rangle -fl(l+1)\left(\frac{1}{x^{2}}+\frac{1}{y^{2}}\right )\\
&\quad\times\langle\phi_{lm}(x)\delta_{lm}(y)\rangle +\frac{l^{2}(l+1)^{2}}{x^{2}y^{2}}\langle\phi_{lm}(x)\phi_{lm}(y)\rangle .
\end{split}
\end{equation}

\section{Derivation of \boldmath{$X_{l}^{2}$} for the analytic cases}
\label{sec:cX}
Here we provide details concerning the calculation of the two analytic examples discussed in section \ref{sec:analytic}.
For the first example, i.e. $\xi\propto 1/\lvert\vx - \vy\rvert$, we have 
\begin{equation} 
\xi(\lvert\va -\va^{\prime}\rvert) = \frac{\sigma_{0}^{2}x_{0}}{\lvert\va -\va^{\prime}\rvert}
= \sigma_{0}^{2}x_{0}\sum_{l^{\prime}} \frac{a^{\prime l^{\prime}}}{a^{l^{\prime}+1}}\frac{4\pi}{2l^{\prime}+1}
\sum_{m^{\prime}}Y_{l^{\prime}m^{\prime}}(\hva)Y^{*}_{l^{\prime}m^{\prime}}(\hva^{\prime}).
\end{equation}
Substituting the above into eq. \eqref{eq:dlm} immediately yields
\begin{equation}
\label{eq:equiv}
\langle\delta_{lm}(a)\delta_{lm}(a^{\prime})\rangle =\frac{4\pi\sigma_{0}^{2}x_{0}}{2l+1}\frac{a^{\prime l}}{a^{l+1}}.
\end{equation}
Using this result in eq. \eqref{eq:cxx} and replacing the double integral over the whole plane $(a,a^{\prime})$ with
twice the integral over the half-plane $(a^{\prime} < a)$, one obtaines
\begin{equation}
X_l^{2} = \left (\frac{f}{2l+1}\right )^{2}\frac{4\pi\sigma_{0}^{2}x_{0}}{2l+1}\frac{2}{(2l-3)(2l-1)}\frac{1}{R^{2l-3}},\quad l\geq 2.
\end{equation}
To deal with the second case, $\xi\propto 1/\lvert\vx - \vy\rvert^{3}$, we write 
\begin{equation}
\frac{1}{\lvert\va -\va^{\prime}\rvert^3} = \frac{1}{aa^{\prime}}\frac{\partial}{\partial\mu}\frac{1}{\lvert\va -\va^{\prime}\rvert}
= \frac{1}{a^{3}}\sum_{l^{\prime}\geq 0}\left (\frac{a^{\prime}}{a}\right )^{l^{\prime}}\frac{\dd\mathcal{P}_{l^{\prime}+1}(\mu )}{\dd\mu},
\end{equation}
where the last step explicitly assumes $a^{\prime} < a$ and $\mu = \hva\cdot\hva^{\prime}$. Introducing
$\vert\vert \mathcal{P}_{l}\vert\vert$ as the usual norm over the interval $\lbrack -1,1\rbrack$ and using the well-known
recurrence relation
\begin{equation}
\frac{\dd\mathcal{P}_{l+1}(\mu)}{\dd\mu} = \frac{2\mathcal{P}_{l}}{\vert\vert\mathcal{P}_{l}\vert\vert^{2}} +
\frac{2\mathcal{P}_{l-2}}{\vert\vert\mathcal{P}_{l-2}\vert\vert^{2}} +
\frac{2\mathcal{P}_{l-4}}{\vert\vert\mathcal{P}_{l-4}\vert\vert^{2}} + \dots ,
\end{equation}
one easily shows that 
\begin{equation}
\begin{split}
\int_{-1}^{1}\dd\mu\mathcal{P}_{l}(\mu)\sum_{l^{\prime}\geq 0}\left (\frac{a^{\prime}}{a}\right )^{l^{\prime}}
\frac{\dd\mathcal{P}_{l^{\prime}+1}(\mu )}{\dd\mu} &= \sum_{l^{\prime}\geq 0}\left (\frac{a^{\prime}}{a}
\right )^{l^{\prime}}\int_{-1}^{1}\dd\mu\frac{\dd\mathcal{P}_{l^{\prime}+1}(\mu)}{\dd\mu}\mathcal{P}_{l}(\mu)\\
&= 2\sum_{j\geq 0}\left (\frac{a^{\prime}}{a}\right )^{l+2j} = \frac{2}{1-(a^{\prime}/a)^{2}}\left (\frac{a^{\prime}}{a}\right )^{l}.
\end{split}
\end{equation}
The above identity may then used to derive the counterpart to eq. \eqref{eq:equiv}.

\section{Expected covariance of multipole estimators}
\label{app:var}

To evaluate the integral expression in eq. \eqref{eq:detect2}, we follow the lines of \cite{Peeb80} and start from the estimator of the
{\sspa} power spectrum multipole $P_{l}(k)$ (omitting any additional superscripts for brevity of notation)
\begin{equation}
\hat{P}_{l}(\vk ) = \frac{(2l+1)V}{4\pi}\int\dd\Omega_{\vk}\mathcal{P}_{l}\left (\mu_{\vk}\right )\delta_{\vk}\delta_{-\vk},
\label{eq:appd1}
\end{equation}
where
\begin{equation}
\delta_{\vk} = \frac{1}{\overline{n}V}\sum\limits_{i}\left (N_{i} - \langle N_{i}\rangle\right )\mathrm{e}^{i\vk\vr_{i}},
\end{equation}
$\mu_{\vk}=k_{3}/k$ is the cosine of the angle between the line of sight and the wave vector, and the volume V has been divided
into infinitesimal cells with occupation numbers $N_{i}=0,1$, i.e. the probability of finding more than one galaxy in a cell is
an infinitesimal of higher order. Using $\langle N_{i}^{2}\rangle = \langle N_{i}\rangle = \overline{n}\delta V_{i}$
and $\langle N_{i}N_{j}\rangle = \overline{n}^{2}\delta V_{i}\delta V_{j}(1+\xi_{ij})$, one immediately verifies that the expectation value
of $\hat{P}_{l}(\vk )$ is given by \cite[cf.][]{Laix1998}
\begin{equation}
\begin{split}
\left\langle\hat{P}_{l}(\vk )\right\rangle &=
\frac{(2l+1)V}{4\pi}\int\dd\Omega_{\vk}\mathcal{P}_{l}\left (\mu_{\vk}\right )\left\langle\delta^{\rm s}_{\vk}\delta^{\rm s}_{-\vk}\right\rangle\\
&= \frac{2l+1}{4\pi\overline{n}^{2}V}\int\dd\Omega_{\vk}\mathcal{P}_{l}\left (\mu_{\vk}\right )\sum\limits_{i,j}
\left\langle (N_{i} - \langle N_{i}\rangle )(N_{j} - \langle N_{j}\rangle )\right\rangle\mathrm{e}^{i\vk (\vr_{i}-\vr_{j})}\\
&= \frac{2l+1}{4\pi\overline{n}^{2}V}\int\dd\Omega_{\vk}\mathcal{P}_{l}\left (\mu_{\vk}\right )\left (\sum\limits_{i\neq j}\overline{n}^{2}
\delta V_{i}\delta V_{j}\xi_{ij}\mathrm{e}^{i\vk (\vr_{i}-\vr_{j})} + \sum\limits_{i}\overline{n}\delta V_{i}\right ).
\end{split}
\label{eq:appd2}
\end{equation}
Restricting the choice of $l$ to positive even numbers, the continuum limit of eq. \eqref{eq:appd2} further yields
\begin{equation}
\left\langle\hat{P}_{l}(\vk )\right\rangle =
\frac{2l+1}{4\pi}\int\dd\Omega_{\vk}\mathcal{P}_{l}\left (\mu_{\vk}\right )\left\lbrack P(\vk ) +
\overline{n}^{-1}\right\rbrack = P_{l}(k).
\label{eq:appd3}
\end{equation}
Thus $\hat{P}_{2}(\vk )$ and $\hat{P}_{4}(\vk )$ are unbiased estimators of the quadrupole and hexadecapole of the {\sspa} power
spectrum $P(\vk)$. To compute the covariance
\begin{equation}
\begin{split}
\left\langle\Delta\hat{P}_{l}(\vk)\Delta\hat{P}_{l}(\vk^{\prime})\right\rangle &=
\left\langle\left\lbrack\hat{P}_{l}(\vk )-\left\langle\hat{P}_{l}(\vk )\right\rangle\right\rbrack\left\lbrack
\hat{P}_{l}(\vk^{\prime})-\left\langle\hat{P}_{l}(\vk^{\prime})\right\rangle\right\rbrack\right\rangle\\
&= \left\langle\hat{P}_{l}(\vk )\hat{P}_{l}(\vk^{\prime})\right\rangle - \left\langle
\hat{P}_{l}(\vk )\right\rangle\left\langle\hat{P}_{l}(\vk^{\prime})\right\rangle,
\end{split}
\label{eq:appd4}
\end{equation}
we consider the expression
\begin{equation}
\left\langle\hat{P}_{l}(\vk )\hat{P}_{l}(\vk^{\prime})\right\rangle = \left\lbrack\frac{(2l+1)V}{4\pi}\right\rbrack^{2}
\int\dd\Omega_{\vk}\int\dd\Omega_{\vk^{\prime}}\mathcal{P}_{l}\left (\mu_{\vk}\right )\mathcal{P}_{l}\left (\mu_{\vk^{\prime}}
\right )\left\langle\delta_{\vk}\delta_{-\vk}\delta_{\vk^{\prime}}\delta_{-\vk^{\prime}}\right\rangle
\label{eq:appd5}
\end{equation}
with
\begin{equation}
\begin{split}
\left\langle\delta_{\vk}\delta_{-\vk}\delta_{\vk^{\prime}}\delta_{-\vk^{\prime}}\right\rangle &= \frac{1}{\left (\overline{n}V
\right )^{4}}\sum\limits_{i,j}\sum\limits_{i^{\prime},j^{\prime}}\mathrm{e}^{i\vk (\vr_{i}-\vr_{j})}
\mathrm{e}^{i\vk^{\prime}(\vr_{i^{\prime}}-\vr_{j^{\prime}})}\\
&{ } \times\left\langle (N_{i} - \langle N_{i}\rangle )(N_{j} - \langle N_{j}\rangle )(N_{i^{\prime}} - \langle N_{i^{\prime}}
\rangle )(N_{j^{\prime}} - \langle N_{j^{\prime}}\rangle )\right\rangle.
\end{split}
\label{eq:appd6}
\end{equation}
As before, eq. \eqref{eq:appd6} may easily be recast in terms of sums over correlation functions. Assuming that the density contrast is
a Gaussian random field, the case of no overlapping indices leads to
\begin{equation}
\begin{split}
&\left\langle (N_{i} - \langle N_{i}\rangle )(N_{j} - \langle N_{j}\rangle )(N_{i^{\prime}} - \langle N_{i^{\prime}}
\rangle )(N_{j^{\prime}} - \langle N_{j^{\prime}}\rangle )\right\rangle\\
&= \overline{n}^{4}\delta V_{i}\delta V_{j}
\delta V_{i^{\prime}}\delta V_{j^{\prime}}\left (\xi_{ij}\xi_{i^{\prime}j^{\prime}} + \xi_{ii^{\prime}}\xi_{jj^{\prime}} + \xi_{ij^{\prime}}\xi_{ji^{\prime}}\right )
\end{split}
\end{equation}
which gives the usual result due to cosmic variance. All remaining configurations contribute to shot noise and can be evaluated in a
similar fashion \cite{Peeb80,Meiksin1999}. Putting everything together and taking the continuum limit, one eventually obtains to leading
order in $(\overline{n}V)^{-1}$
\begin{equation}
\begin{split}
\left\langle\Delta\hat{P}_{l}(\vk)\Delta\hat{P}_{l}(\vk^{\prime})\right\rangle &= \left (\frac{2l+1}{4\pi}\right )^{2}\int\dd\Omega_{\vk}
\int\dd\Omega_{\vk^{\prime}}\mathcal{P}_{l}\left (\mu_{\vk}\right )\mathcal{P}_{l}\left (\mu_{\vk^{\prime}}\right )\\
&\times\frac{(2\pi)^{3}}{V}\left\lbrack\left (1+f\mu_{\vk}^{2}\right )^{2}P(k)+\overline{n}^{-1}\right\rbrack^{2}\left\lbrack
\delta_{D}(\vk -\vk^{\prime}) + \delta_{D}(\vk +\vk^{\prime})\right\rbrack,
\end{split}
\end{equation}
where we have used $\mu_{\vk}^{2}=\mu_{-\vk}^{2}$ and the Kaiser relation $P(\vk )=(1+f\mu_{\vk}^{2})^{2}P(k)$ for $b=1$. Remembering
that $\mathcal{P}_{l}(\mu_{-\vk})=(-1)^{l}\mathcal{P}_{l}(\mu_{\vk})$ and carrying out the integral over $\dd\Omega_{\vk^{\prime}}$,
one arrives at
\begin{equation}
\begin{split}
\left\langle\Delta\hat{P}_{l}(\vk)\Delta\hat{P}_{l}(\vk^{\prime})\right\rangle &=
\frac{(2l+1)^{2}}{4\pi V}\frac{(2\pi)^{3}}{k^{\prime 2}}\delta_{D}(k -k^{\prime})\int_{-1}^{1}\dd\mu_{\vk}\mathcal{P}_{l}^{2}
\left (\mu_{\vk}\right )\left\lbrack P(\vk )+\overline{n}^{-1}\right\rbrack^{2}\\
&= \frac{(2l+1)^{2}}{4\pi V}\frac{(2\pi)^{3}}{k^{\prime 2}}\delta_{D}(k -k^{\prime})\left\lbrack\mathcal{C}_{l4}P^{2}(k) +
2\mathcal{C}_{l2}P(k)\overline{n}^{-1} + \mathcal{C}_{l0}\overline{n}^{-2}\right\rbrack,
\end{split}
\end{equation}
with $l$ assumed as even and $\mathcal{C}_{ll^{\prime}}$ defined in eq. \eqref{eq:detect3b}. Inserting this result into
eq. \eqref{eq:detect2} and integrating over $\dd^{3}k^{\prime}$, one finally ends up with the expression given by eq \eqref{eq:detect3}.

\bibliography{RSD.bib}

\providecommand{\href}[2]{#2}\begingroup\raggedright\begin{thebibliography}{10}

\bibitem{Einstein}
A.~{Einstein}, {\it {Zum kosmologischen Problem der allgemeinen
  Relativit\"{a}tstheorie}},  {\em Sitzungsberichte der Preu{\ss}ischen
  Akademie der Wissenschaften zu Berlin} {\bf 16 April 1931} (1931) 235--237.

\bibitem{Milne}
E.~A. {Milne}, {\it {World-Structure and the Expansion of the Universe. Mit 6
  Abbildungen.}},  {\em Zeitschrift f\"{u}r Astrophysik} {\bf 6} (1933) 1.

\bibitem{Milne35}
E.~A. {Milne}, {\em {Relativity, gravitation and world-structure}}.
\newblock {Oxford, The Clarendon press}, 1935.

\bibitem{Peeb80}
P.~J.~E. {Peebles}, {\em {The large-scale structure of the universe}}.
\newblock {Princeton University Press}, 1980.

\bibitem{Jackson1972}
J.~C. {Jackson}, {\it {A critique of Rees's theory of primordial gravitational
  radiation}},  {\em \mnras} {\bf 156} (1972) 1P.

\bibitem{k87}
N.~{Kaiser}, {\it {Clustering in real space and in redshift space}},  {\em
  \mnras} {\bf 227} (July, 1987) 1--21.

\bibitem{Fisher95b}
K.~B. {Fisher}, O.~{Lahav}, Y.~{Hoffman}, D.~{Lynden-Bell}, and S.~{Zaroubi},
  {\it {Wiener reconstruction of density, velocity and potential fields from
  all-sky galaxy redshift surveys}},  {\em \mnras} {\bf 272} (Feb., 1995)
  885--908, [\href{http://xxx.lanl.gov/abs/astro-ph/9406009}{{\tt
  astro-ph/9406009}}].

\bibitem{Hamilton1998}
A.~J.~S. {Hamilton}, {\it {Linear Redshift Distortions: a Review}},  in {\em
  The Evolving Universe} (D.~{Hamilton}, ed.), vol.~231 of {\em Astrophysics
  and Space Science Library}, p.~185, 1998.

\bibitem{Scoccimarro2004}
R.~{Scoccimarro}, {\it {Redshift-space distortions, pairwise velocities, and
  nonlinearities}},  {\em \prd} {\bf 70} (Oct., 2004) 083007,
  [\href{http://xxx.lanl.gov/abs/astro-ph/0407214}{{\tt astro-ph/0407214}}].

\bibitem{chisari}
N.~E. {Chisari} and M.~{Zaldarriaga}, {\it {Connection between Newtonian
  simulations and general relativity}},  {\em \prd} {\bf 83} (June, 2011)
  123505, [\href{http://xxx.lanl.gov/abs/1101.3555}{{\tt arXiv:1101.3555}}].

\bibitem{wald12}
S.~R. {Green} and R.~M. {Wald}, {\it {Newtonian and relativistic cosmologies}},
   {\em \prd} {\bf 85} (Mar., 2012) 063512,
  [\href{http://xxx.lanl.gov/abs/1111.2997}{{\tt arXiv:1111.2997}}].

\bibitem{SW}
R.~K. {Sachs} and A.~M. {Wolfe}, {\it {Perturbations of a Cosmological Model
  and Angular Variations of the Microwave Background}},  {\em \apj} {\bf 147}
  (Jan., 1967) 73.

\bibitem{BT}
J.~{Binney} and S.~{Tremaine}, {\em {Galactic Dynamics: Second Edition}}.
\newblock Princeton University Press, 2008.

\bibitem{ND94}
A.~{Nusser} and M.~{Davis}, {\it {On the prediction of velocity fields from
  redshift space galaxy samples}},  {\em \apjl} {\bf 421} (Jan., 1994) L1--L4,
  [\href{http://xxx.lanl.gov/abs/astro-ph/9309009}{{\tt astro-ph/9309009}}].

\bibitem{wmap7}
D.~{Larson}, J.~{Dunkley}, G.~{Hinshaw}, E.~{Komatsu}, M.~R. {Nolta}, C.~L.
  {Bennett}, B.~{Gold}, M.~{Halpern}, and {et al.}, {\it {Seven-year Wilkinson
  Microwave Anisotropy Probe (WMAP) Observations: Power Spectra and
  WMAP-derived Parameters}},  {\em \apjs} {\bf 192} (Feb., 2011) 16,
  [\href{http://xxx.lanl.gov/abs/1001.4635}{{\tt arXiv:1001.4635}}].

\bibitem{EH98}
D.~J. {Eisenstein} and W.~{Hu}, {\it {Baryonic Features in the Matter Transfer
  Function}},  {\em \apj} {\bf 496} (Mar., 1998) 605--+,
  [\href{http://xxx.lanl.gov/abs/astro-ph/9709112}{{\tt astro-ph/9709112}}].

\bibitem{vsh}
R.~G. {Barrera}, G.~A. {Estevez}, and J.~{Giraldo}, {\it {Vector spherical
  harmonics and their application to magnetostatics}},  {\em European Journal
  of Physics} {\bf 6} (Oct., 1985) 287--294.

\bibitem{Hamilton1992}
A.~J.~S. {Hamilton}, {\it {Measuring Omega and the real correlation function
  from the redshift correlation function}},  {\em \apjl} {\bf 385} (Jan., 1992)
  L5--L8.

\bibitem{bigboss2011}
D.~{Schlegel}, F.~{Abdalla}, T.~{Abraham}, C.~{Ahn}, C.~{Allende Prieto},
  J.~{Annis}, E.~{Aubourg}, M.~. {Azzaro}, and et~al., {\it {The BigBOSS
  Experiment}},  {\em ArXiv e-prints} (June, 2011)
  [\href{http://xxx.lanl.gov/abs/1106.1706}{{\tt arXiv:1106.1706}}].

\bibitem{euclid09}
R.~{Laureijs}, {\it {Euclid Assessment Study Report for the ESA Cosmic
  Visions}},  {\em ArXiv:0912.0914} (Dec., 2009)
  [\href{http://xxx.lanl.gov/abs/0912.0914}{{\tt arXiv:0912.0914}}].

\bibitem{Amendola2013}
L.~{Amendola}, S.~{Appleby}, D.~{Bacon}, T.~{Baker}, M.~{Baldi}, N.~{Bartolo},
  A.~{Blanchard}, C.~{Bonvin}, S.~{Borgani}, and {et al.}, {\it {Cosmology and
  Fundamental Physics with the Euclid Satellite}},  {\em Living Reviews in
  Relativity} {\bf 16} (Sept., 2013) 6,
  [\href{http://xxx.lanl.gov/abs/1206.1225}{{\tt arXiv:1206.1225}}].

\bibitem{Hui2007}
L.~{Hui}, E.~{Gazta{\~n}aga}, and M.~{Loverde}, {\it {Anisotropic magnification
  distortion of the 3D galaxy correlation. I. Real space}},  {\em \prd} {\bf
  76} (Nov., 2007) 103502, [\href{http://xxx.lanl.gov/abs/0706.1071}{{\tt
  arXiv:0706.1071}}].

\bibitem{Hui2008}
L.~{Hui}, E.~{Gazta{\~n}aga}, and M.~{Loverde}, {\it {Anisotropic magnification
  distortion of the 3D galaxy correlation. II. Fourier and redshift space}},
  {\em \prd} {\bf 77} (Mar., 2008) 063526,
  [\href{http://xxx.lanl.gov/abs/0710.4191}{{\tt arXiv:0710.4191}}].

\bibitem{planck13}
{Planck collaboration}, P.~A.~R. {Ade}, N.~{Aghanim}, C.~{Armitage-Caplan},
  M.~{Arnaud}, M.~{Ashdown}, F.~{Atrio-Barandela}, J.~{Aumont},
  C.~{Baccigalupi}, A.~J. {Banday}, and et~al., {\it {Planck 2013 results. XV.
  CMB power spectra and likelihood}},  {\em ArXiv:1303.5075} (Mar., 2013)
  [\href{http://xxx.lanl.gov/abs/1303.5075}{{\tt arXiv:1303.5075}}].

\bibitem{bossbao}
H.~{Guo}, I.~{Zehavi}, Z.~{Zheng}, D.~H. {Weinberg}, A.~A. {Berlind},
  M.~{Blanton}, Y.~{Chen}, D.~J. {Eisenstein}, and {et al.}, {\it {The
  Clustering of Galaxies in the SDSS-III Baryon Oscillation Spectroscopic
  Survey: Luminosity and Color Dependence and Redshift Evolution}},  {\em \apj}
  {\bf 767} (Apr., 2013) 122, [\href{http://xxx.lanl.gov/abs/1212.1211}{{\tt
  arXiv:1212.1211}}].

\bibitem{planckanomalies}
{Planck Collaboration}, P.~A.~R. {Ade}, N.~{Aghanim}, C.~{Armitage-Caplan},
  M.~{Arnaud}, M.~{Ashdown}, F.~{Atrio-Barandela}, J.~{Aumont},
  C.~{Baccigalupi}, A.~J. {Banday}, and et~al., {\it {Planck 2013 results. XVI.
  Cosmological parameters}},  {\em ArXiv:1303.5076} (Mar., 2013)
  [\href{http://xxx.lanl.gov/abs/1303.5076}{{\tt arXiv:1303.5076}}].

\bibitem{Laix1998}
A.~A. {de Laix} and G.~{Starkman}, {\it {Sensitivity of Redshift Distortion
  Measurements to Cosmological Parameters}},  {\em \apj} {\bf 501} (July, 1998)
  427, [\href{http://xxx.lanl.gov/abs/astro-ph/9707008}{{\tt
  astro-ph/9707008}}].

\bibitem{Meiksin1999}
A.~{Meiksin} and M.~{White}, {\it {The growth of correlations in the matter
  power spectrum}},  {\em \mnras} {\bf 308} (Oct., 1999) 1179--1184,
  [\href{http://xxx.lanl.gov/abs/astro-ph/9812129}{{\tt astro-ph/9812129}}].

\end{thebibliography}\endgroup
\end{document}